# Recent results on the low mass dark matter WIMP controversy: 2011

David B. Cline
Astroparticle Physics Group
UCLA Physics & Astronomy

**Abstract**

We review the confused situation concerning evidence for low-mass WIMPs. In the past one half year there have been new results concerning the existence of WIMPs at low mass including the new XENON 100, 100-day data, additional CDMS results, the publication of annual variation data from LVD and Borexino and new CoGeNT data. Along with the $S_2$ analyses of the XENON 10 data we provide an overview of this situation. We discuss new results from 2011 here. We also discuss the origin of annual variations of signals in underground laboratories.

This article is meant to be an update of recent experimental results. It is not a critical comparison of the claims of various experimental groups. Such critiques are made in public conferences and meetings. There is currently an intense discussion being carried out about the low mass WIMP region with many different viewpoints. We have little to say about this situation except that the scientific method usually insures the correct results will eventually surface.

## 1. Current state of world data 2011: Overview of new results

The DAMA/LIBRA results have been extensively studied by G. Gelmini [1] and colleagues. They have fit all the data and find two WIMP mass regions as shown in Figure 1. We study here the low mass region (~10GeV) which has a lower probability in their fit. The higher mass solution is ruled out by a large factor by 10 or more by direct dark matter search experiments. [5 experiments] [see references 2,3]. Note that the low mass region has a lower probability in Figure 1.

These results and the possible WIMP signal from the CoGeNT experiment along with the current constraints taken on face value from CDMSII, XENON10, and XENON100 data are shown in Figure 2 [4]. The constraints on the low mass region (~7GeV/c$^2$) have been called into question recently.

We note that CRESST has not published their claims in a peer-reviewed journal yet but seems to be making some claim for a signal above background in the detector. In addition the DAMA result is ruled out by both the XENON 100 and special CDMS data (to be discussed later). Thus there are six experiments to be discussed here that impact the low mass region:

(1) CDMS low energy data [Figure 8]
(2) XENON 100 (100 days) [Figure 12]
(3) An analysis of the XENON 10 data using only the $S_2$ signal, independent of the so-called $L_{eff}$ parameter (to be discussed later) [Figure 7]
(4) A new measurement of $L_{eff}$ by the Columbia group [Figure 11]
(5) New results from the SIMPLE experiment [Figure 9]

(6) A critique of low mass results by the CoGeNT group [Figure 10]

Note that the members of XENON 10 and XENON 100 groups are very different. The all $S_2$ analysis is independent of the normal ($S_1$, $S_2$, $L_{eff}$) analysis of XENON 100. These are then two totally independent experiments. For the recent CoGeNT results see C.E. Aslseth et al, Phys Letters **106**, 2011, 131301. The authors of CoGeNT, CRESST, and DAMA indicate that the three experiments show the evidence for low-mass WIMPs. However the DAMA results are excluded by two CDMS experiments. Therefore three to four independent experiments exclude the low WIMP mass hypothesis. We will discuss these more below.

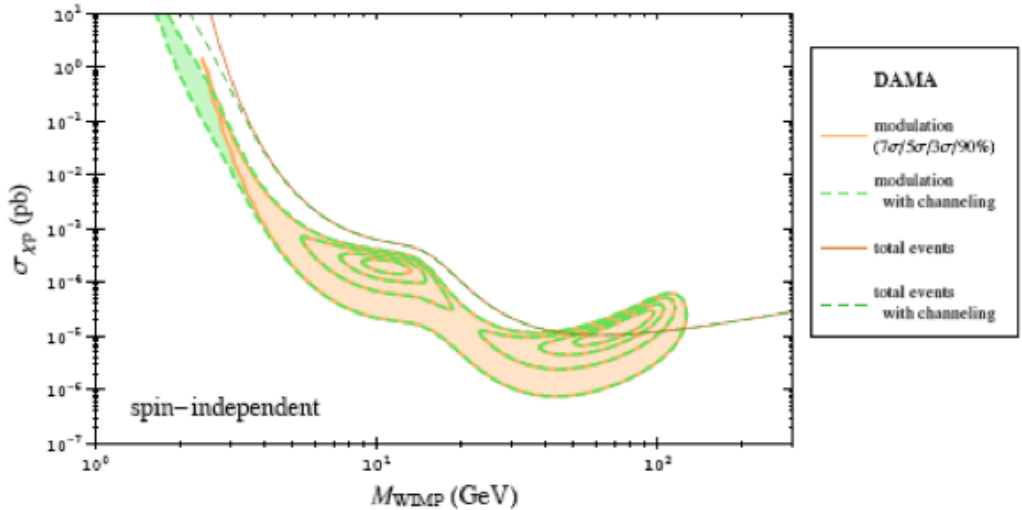

Figure 1. WIMP masses and spin-independent (SI) cross-sections compatible with the DAMA modulation signal and total number of events, determined with (dashed green) and without (solid orange) the channeling effect included. The largest channeling fractions shown in Figure 1 (taken from Ref. [3]) are used here for the channeling case. Comparing the cases with or without channeling, we find negligible difference in the DAMA modulation regions at the 90%, 3σ, and 5σ levels; only the 7σ contours differ and only for WIMP masses below 4 GeV. The lower and higher mass DAMA regions correspond to parameters where the modulation signals arise from scattering predominantly off of NA and I, from Reference 4.

**CoGeNT and DAMA (from Reference 4 for references)**

1) The CoGeNT experiment uses a small Ge crystal to search for a coherent interaction than just experiment used reaction neutrinos to search for antineutrino coherent scattering. They made impressive gains but the low energy background obscured the effect. The detector was sent underground to Soudon and a search was made of coherent WIMP scattering. A similar signal on background was observed (see Juan Collar's talk at DM 2010 website, http://www.physics.ucla.edu/hep/dm10).[1]
2) The DAMA /Libra experiment is well known. Reference 4 gives many of the original references.

---

[1] We thank Dr. Collar for discussions.

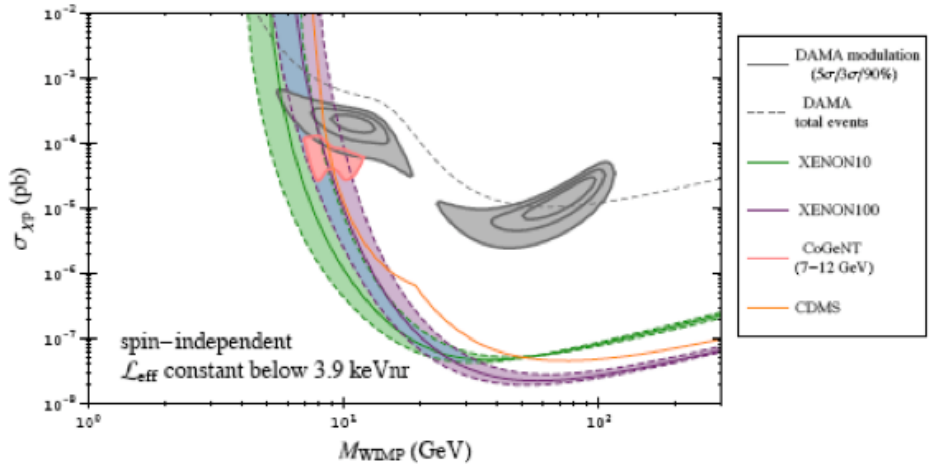

Figure 2. XENON10 (green) and XENON100 (purple) 90% C.L. constraints for a constant $L_{eff}$ at recoil energies below 3.9 KeVnr. The solid curves are the constraints using the central values of $L_{eff}$ as described in the text; dashed curves and lighter filled regions indicate how these 90% constraints vary with the $1\sigma$ uncertainties in $L_{eff}$. The blue region indicates an overlap between the XENON10 (green) and XENON100 (purple) $1\sigma$ regions. Also shown are the CDMS constraint (orange curve), DAMA modulation compatible regions (gray contours/region), and the CoGeNT 7-12 GeV region (pink contour/region). The lower and higher mass DAMA regions correspond to parameters where the modulation signals arise from scattering predominantly off of Na and I, from Reference 4, see also References 2 & 3.

## 2. A new look at the DAMA effect

At a recent meeting in Rapid City, South Dakota, USA, J. Ralston (private communication) made a plot of the annual variation from the ICARUS using ICARUS data from the UCLA website. We show this plot in Figure 3. While this is not a fit, it is strange how well the neutron data fits the DAMA results (see John Ralston, arXiv 1006.5255v 28 June 2010). The interaction of low energy neutrons in either Na or I has many resonances. The effect of these resonances should be studied to see if they could affect the DAMA signal.

## 3. Response of a detector to low mass WIMPs: liquid noble gas detectors

At first sight one would assume that the only important quantity is the mass of the target nucleus. This is only partly true. It is the sensitivity to the partially ionized nucleus that moves a short distance. Consider the liquid noble gas detectors Xe, Ar, etc. After a collision with a WIMP two signals can be observed: scintillation light and electrons. In the ZEPLIN II detector we discovered using incident neutrons projected into the detector that the scintillation light ($S_1$) and electron emission ($S_2$) behaved very differently near the lowest energy recoil. This is likely due to the fact that to strip an electron from a partially ionized Xe atom is relatively easy. It takes about 14 eV to create a single electron from the Xenon.

We observed in 2005 with the ZEPLIN II detector that even when the $S_1$ signal could not be detected there was still a robust $S_2$ signal.[2] This same signal has been observed in XENON 10 and XENON 100. This can only be due to the electrons that are emitted from the moving Xe atom in the liquid Xenon medium.

In part the $S_2$ signal is due to the internal amplification of even a single electron to 25 photo electrons in the two-phase detector. This effect can be seen in the other detectors in Figure 7 using NaI and Ge (and other materials) don't have such an internal amplification system. The electron signal gives the Xenon detector a much lower threshold than is normally expected.

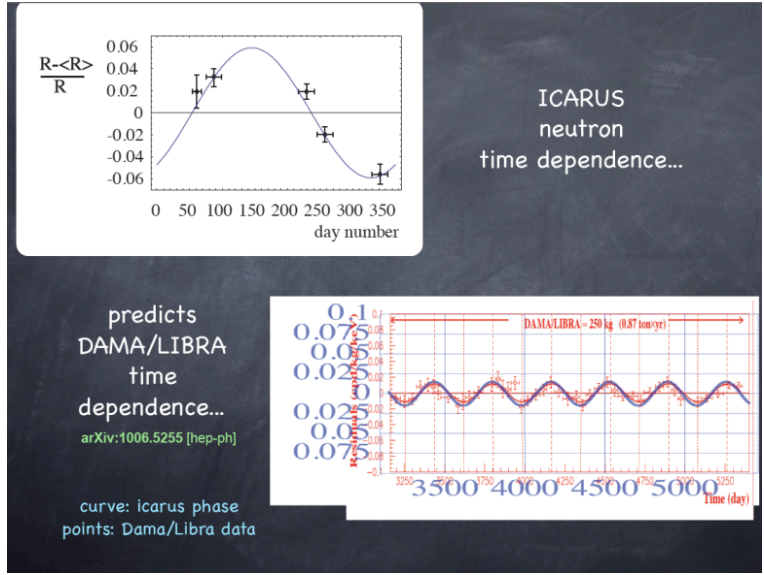

Figure 3. The results from a study by J. Ralston comparing neutron data at the LNGS and the DAMA result.

![Figure 4 image]

Figure 4. Recent study of backgrounds in DAMA.

---

[2] Private information inside the UCLA ZEPLIN II group.

In addition to the study by Ralston there has been a recent very careful study of the single rate in DAMA compared to the expectations from radioactive processes. This is shown in Figure 4. The conclusion is that there is no evidence for WIMP production in the bulk of the data.

**4. The use of the $S_2/S_1$ and $S_2$ methods in two-phase detectors for the low mass regions**

From the start of the concept of the $S_2/S_1$ discrimination in the 1990s, we knew that there would be problems when $S_1$ got very small or near threshold for the design of ZEPLIN II [6]. An internal analysis was made by DBC of the ZEPLIN II data using only $S_2$ data (when $S_1$ was very small). This turned out not to help the sensitivity for ZEPLIN II because of the large PMTs that did not allow a precise fiducial volume to be defined and was never published. However for future analyses the $S_2$ method may be useful.

While it may seem that the low mass WIMP region should be probed by lower mass targets (like oxygen or sodium), this is not the case if the two-phase method is used. In this method electrons generated by the recoil nucleus moving through the medium generate free electrons. The application of an electrical field on the detector is used to drift the free electrons into a volume of gas with a higher electrical field that provides an amplification of the signal. This can overcome any advantage that low mass might offer. Very approximately for every free electron produced at the WIMP interaction vertex there will be a 25 $P_e$ ($P_e$ photoelectric) in the gas phase.

This concept was invented to be used for liquid Xenon detectors at UCLA and Torino universities in the mid-1990s and was demonstrated in the ZEPLIN II event first, but is now being used for liquid Argon detectors (WARP, Ar or Ne, etc.) and others [6]. In this paper we only cover the use in liquid Xenon detectors with published data.

**5. The use of the $S_2$ signal**

The key to the use of $S_2$ is to define a small fiducial volume. X and Y positions are measured directly in the PMT structure. The determination of the Z point of the event uses the time spread of the $S_2$ signal produced at different depths in the detector, since the X,Y signal can define a small fiducial volume even in a poorly measured Z position and maintain this small volume.

As stated before, an internal study using $S_2$ data was carried out for the ZEPLIN II data at UCLA. No analysis was done with the data due to the large backgrounds. More recently a very nice study of the use of $S_2$ data to determine the sensitivity of XENON10 data to low mass WIMPs was done by P. Sorensen [7]. We repeat some of this analysis here for completeness and provide results from the recent XENON 10 $S_2$ analysis.

**6. Exclusion limits for low mass wimps**

There is considerable controversy on the WIMP mass region that is excluded by the XENON10 and XENON100 experiments. There are new and unconfirmed claims of more candidate events from the CRESST experiment. Clearly new measurements are needed.

(1) The XENON100 experiment is currently taking data that is "blinded," namely the data will not be studied until the "box is opened". These data should be at least a factor of 2 more sensitive than current data.
(2) New direct measurements of the quenching factor or $L_{eff}$ need to be carried out. There are plans in the XENON100 group to carry out these measurements in the next year, we understand.

However while waiting for these new measurements we can draw some tentative conclusions from the preceding arguments about:

(1) The effect of uncertainty in $L_{eff}$
(2) The use of the $S_2$ method to probe the sensitivity of the Xenon detectors to signals very close to threshold

The work by Savage et al (Reference 4) and the XENON100 paper both address issue 1.

In the now published XENON100 paper that uses 11 days of "unblinded" data the authors state "as shown in Figure 3 over the acceptance is sizeable even at a reduced threshold of 3PE (8.2 KeV in this case) above which a 7 GeV/$c_2$ WIMP at the lower edge of the CoGeNT region would produce about one event with the current exposure."

In this case the forthcoming new XENON100 data would have between 5-10 events in this case. So the new data should resolve the issue.

**7. Recent CDMS data for the low-mass region**

In Figure 7 we show new data from the CDMS experiment that seems to exclude the low-mass region (Ref. 8).

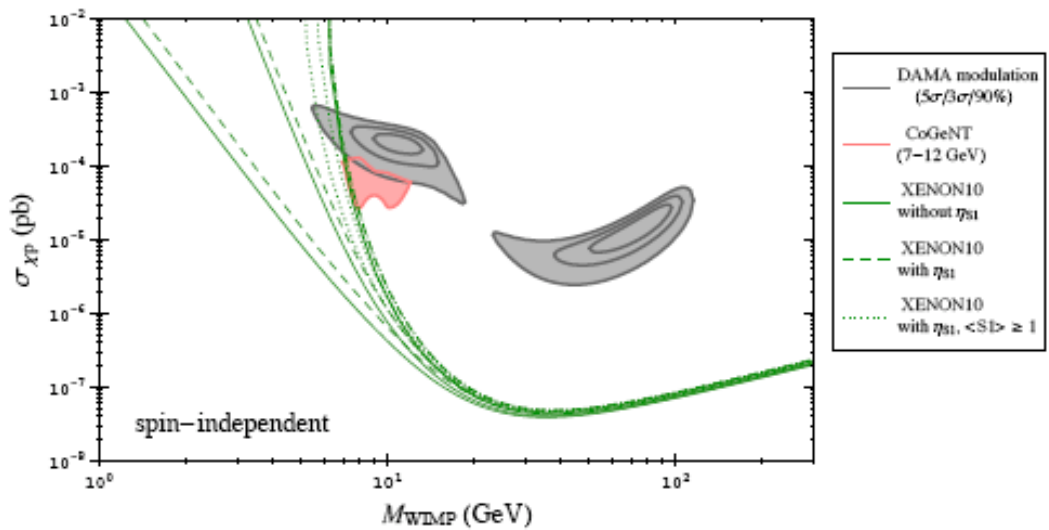

Figure 5 (Reference 4). Exclusion plot of Savage et al.

## 8. The All S$_2$ Analysis

The use of the S$_2$ method to determine the near threshold behavior of L$_{eff}$ (or equivalent) was described in Reference 7. We show the results of this analysis in Figure 6 for this model independent L$_{eff}$ measurement.

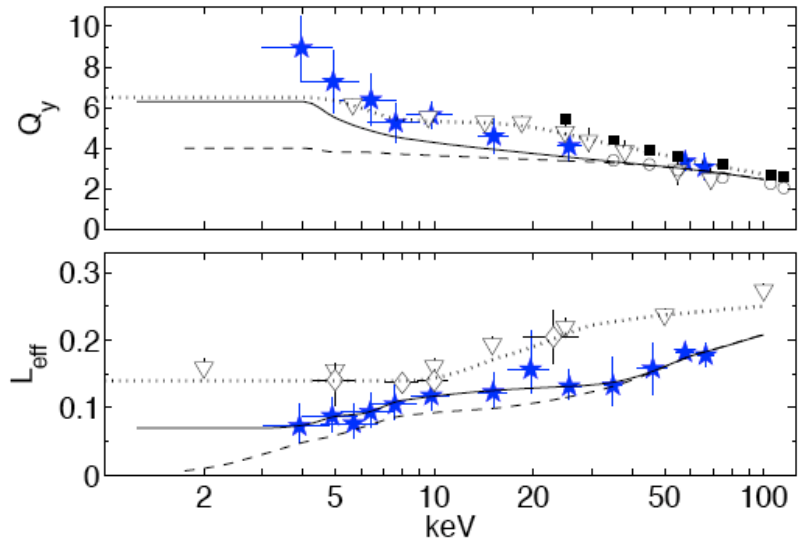

Figure 6. Study of L$_{eff}$ by Sorensen using S$_2$ signal data (see Reference 7). (New method)

The all S$_2$ analysis was first used by Sorensen. The basic idea is to use the S$_2$ data that is due to the electric field in the detector and the amplification on the gas. The x and y coordinates are measured by the PMT signal. Tne z (or upward) coordinate is measured by the dispersion in the electron signal. However, recently the XENON 10 group put out a paper using this method that strongly constrains the low-mass region [Ref. 10]. This group has done a very careful analysis using the S$_2$ signal only. This is essentially independent of the L$_{eff}$ . In Figure 7 we show the recent S$_2$ analysis result of the XENON 10 group (Reference 10). Note that this analysis excludes both the DAMA and CoGeNT results.

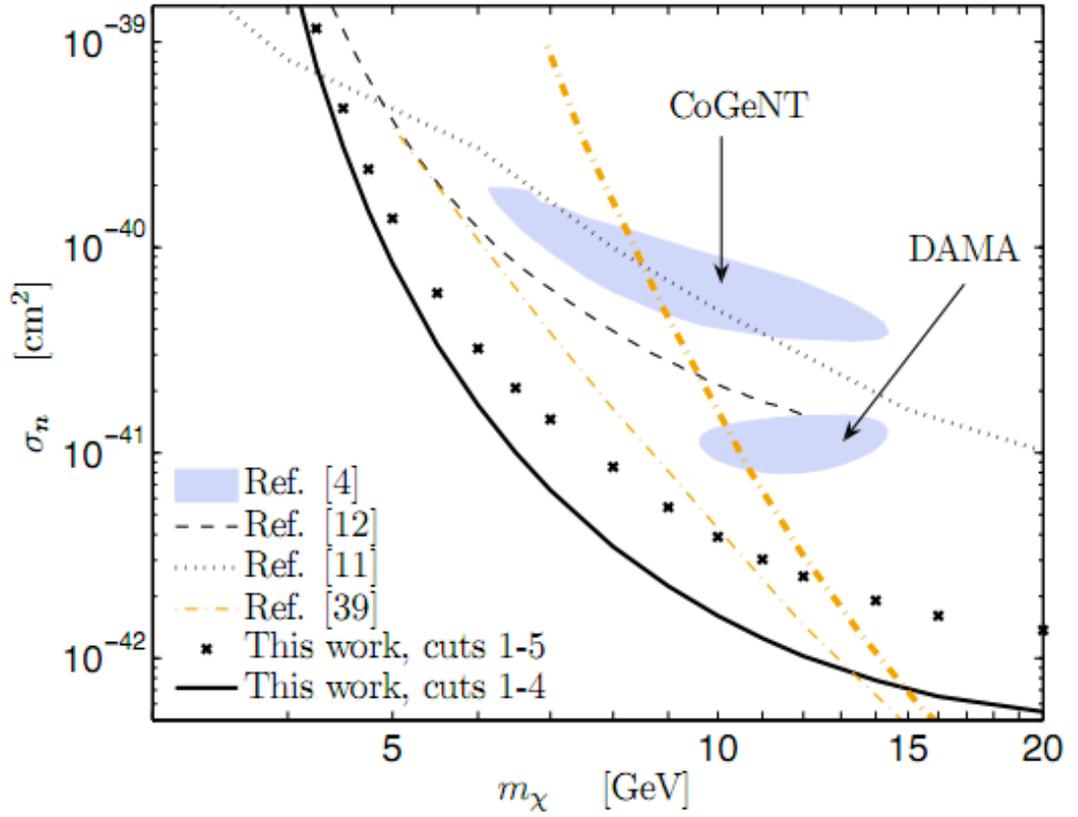

Figure 7 (from Reference 10). Curves indicate 90% C.L. exclusion limits on spin-independent $\sigma_n$ for elastic dark matter scattering, obtained by CDMS (dotted (11) and dashed (12), XENON100 (dash-dot [39]). 99% C.L. allowed regions consistent with the assumption of a positive detection are also shown, for signals from DAMA (with ion channeling) [4], and CoGeNT (assuming 30% exponential background) [4]. See also Figure 9. See Reference 10 for the meaning of these references from the paper.

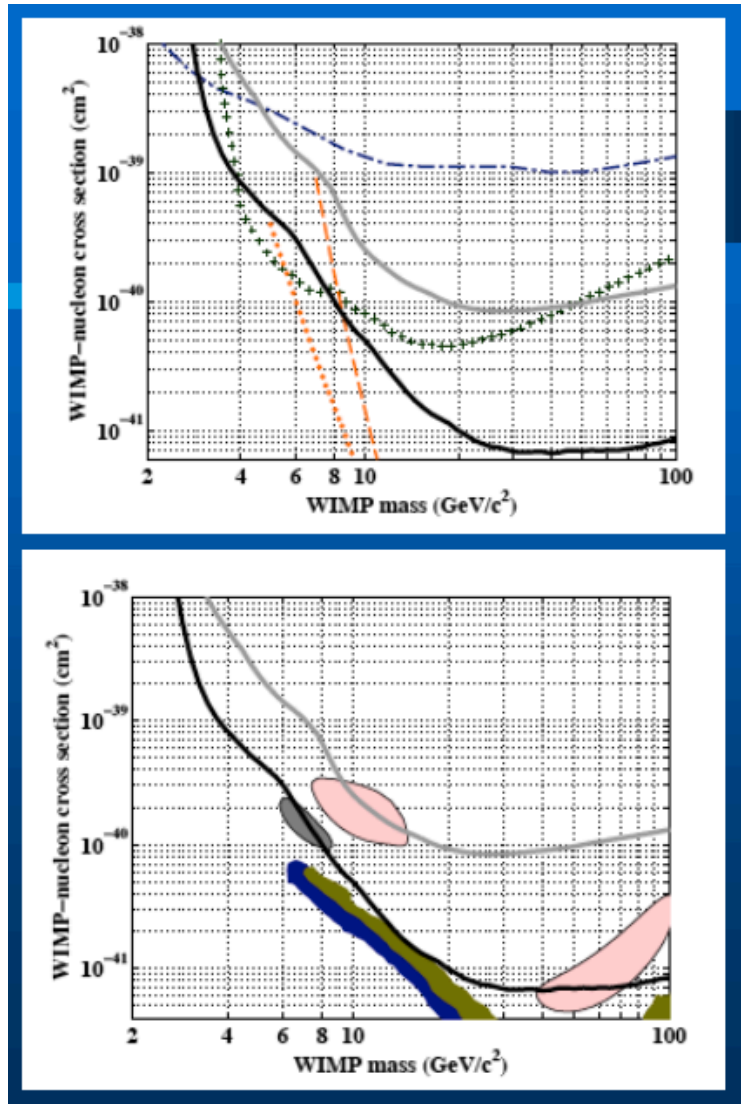

Figure 8. Exclusion plot from CDMS special run at low energy (8).

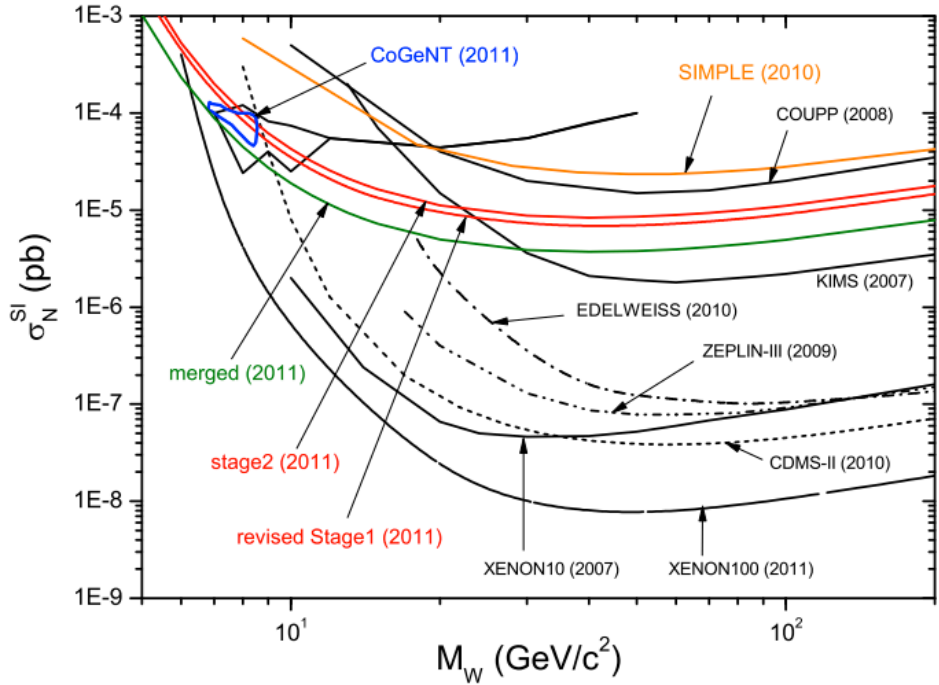

Figure 9. A new result from SIMPLE (Ref. 14), showing various spin-independent contours for SIMPLE, together with those of the leading [21-28] spin-independent search results; shown are both previous and reanalyzed Stage 1 results, Stage 2, and a merging of the two.

The XENON100, Savage et al and the $S_2$ analysis of Sorensen all show a similar pattern of exclusion of the low mass WIMPs. The results of the exclusion plot using this method is shown in Figure 6 [Reference 7].

The current situation in the search for low mass WIMPs is confused. Two experiments give a hint of a signal. However the XENON10 and XENON100 (and CDMS II) seem to rule out much of the phase space and the analyses of Savage et al (Reference 4). Sorensen (Reference 7) is very promising using the $S_2$ signal from the XENON two-phase detectors (XENON10). The new CDMS data seems also to exclude the low-mass region. The new XENON 10 analysis using $S_2$ (Figure 7) seem to confirm this from XENON 100. For a recent summary see Reference 9. There is a new result from SIMPLE (Figure 9) that excludes DAMA as well (Ref. 14).

## 9. An alternative viewpoint on the low mass limits

There have been critical comments on the low mass limits made by Juan Collar.[3] In essence, he points out that the quenching factor for the DAMA data has been overly restrictive and that the liquid Xenon experiments have "over sold" their limits. The point is that for the low mass region a target like Xe is too heavy and Na or Ge or even oxygen would provide a better target (see comments in section 3). Another problem is the lack of direct calibration of the detectors in the low energy record region. All of the Xenon based detectors have been exposed to neutron beams.

---

[3] "A realistic assessment of the sensitivity of XENON 10 and XENON 100 to light mass WIMPs," J. Collar, preprint.

In Figure 10 the crux of this argument is seen. We note that the all S2 analyses (Figure 7) is not show in Figure 10, which is aimed at the recent XENON 100 results (see Figure 7). The region of signal for DAMA in Figure 10 is given by the DAMA group's measured quenching factor. However in Figure 10 a more uncertain quenching factor is used. All of these points need to be addressed and clarified. Data from other detectors such as ZEPLIN III, Edelweiss are also needed.

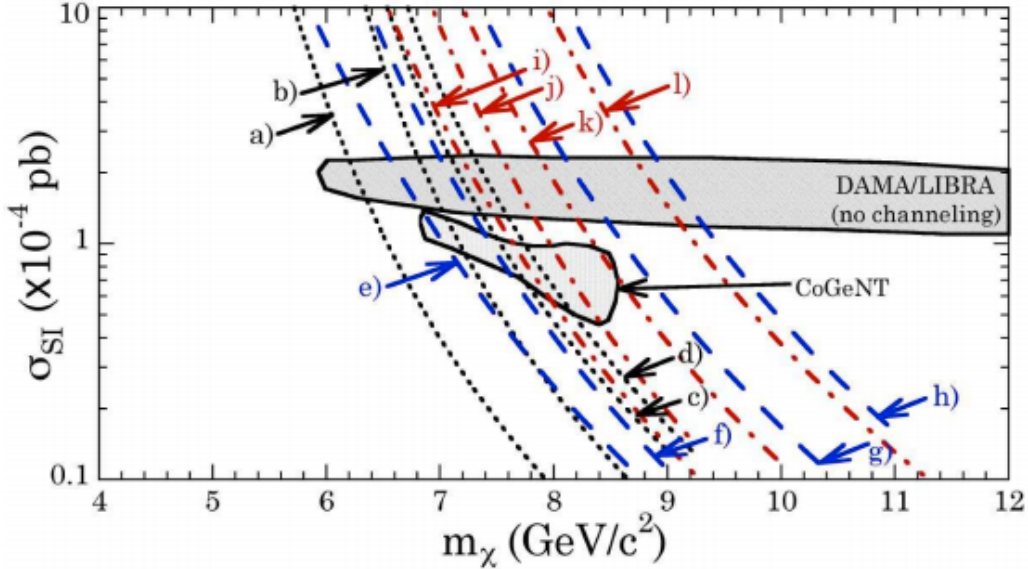

Figure 10. FIG. 1: 90% C.L. XENON100 exclusion contours obtained under the assumptions discussed in the text. Dotted (black) curves correspond to the $L_{eff}$ by Plante et al. [11], dashed (blue) to that by Manzur et al. [14], and dash-dotted (red) to a most recent Monte Carlo-independent $L_{eff}$ determination by ZEPLIN-III [37]. The notation used to describe each case lists number of irreducible recoil events accepted, $L_{eff}$ considered (central value of logarithmic extrapolation or its lower C.L. boundary), and statistics used (see text): a) 1 event, central, Poisson; b) 1 ev., 2σ, Poisson; c) 4 evs., 2σ, Poisson; d) 4evs., 2σ, binomial; e) 1 ev., central, Poisson; f) 1 ev., central, binomial; g) 1 ev., 1σ, Poisson; h) 4 evs., 1σ, binomial; i) 1 ev., central, Poisson; j) 1 ev., central, binomial; k) 4 evs., central, binomial; l) 4 evs., 1σ, binomial. Additional instrumental uncertainties not reflected in this figure are listed in the text.

**10. New results from XENON 100**

The XENON 100 paper using data from 100 days has now been released. There are two new results (Reference 11):

(a) New measurements of $L_{eff}$ by the Columbia group (Figure 11)
(b) A new limit on the low mass WIMP region (Figure 12) from XENON 100.

The new XENON 100 results are shown in Figure 11 and 12. Note that the results in Figure 12 at low mass exclude the CoGeNT and DAMA results.

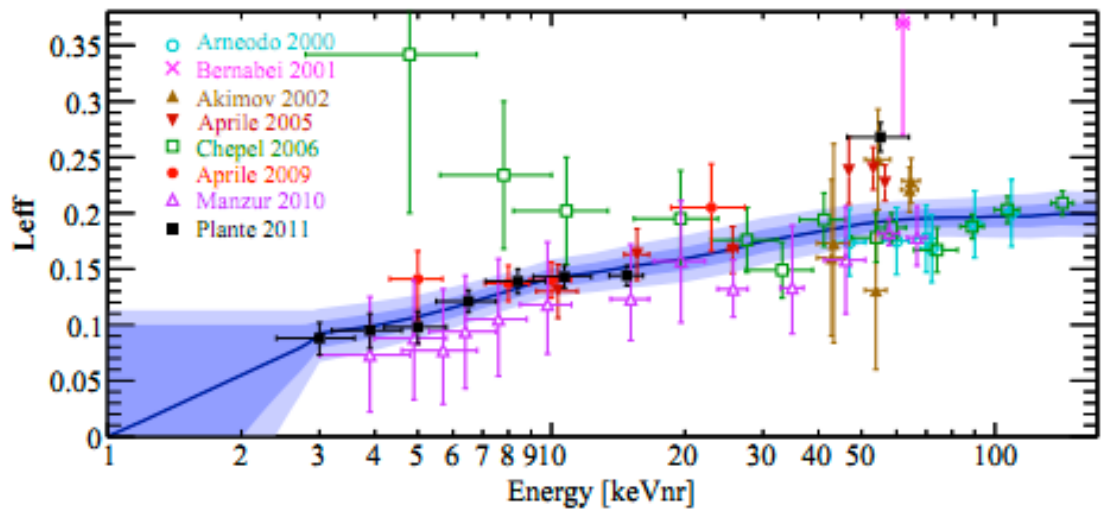

Figure 11. All direct measurements of $L_{eff}$ described by a Gaussian distribution to obtain the mean (solid line) and the uncertainty band (shaded blue, 1σ and 2σ). Below 3 keV$_{nr}$, where no direct measurements exist, the trend is logarithmically extrapolated to $L_{eff}$ = 0 at 1 keV$_{nr}$.

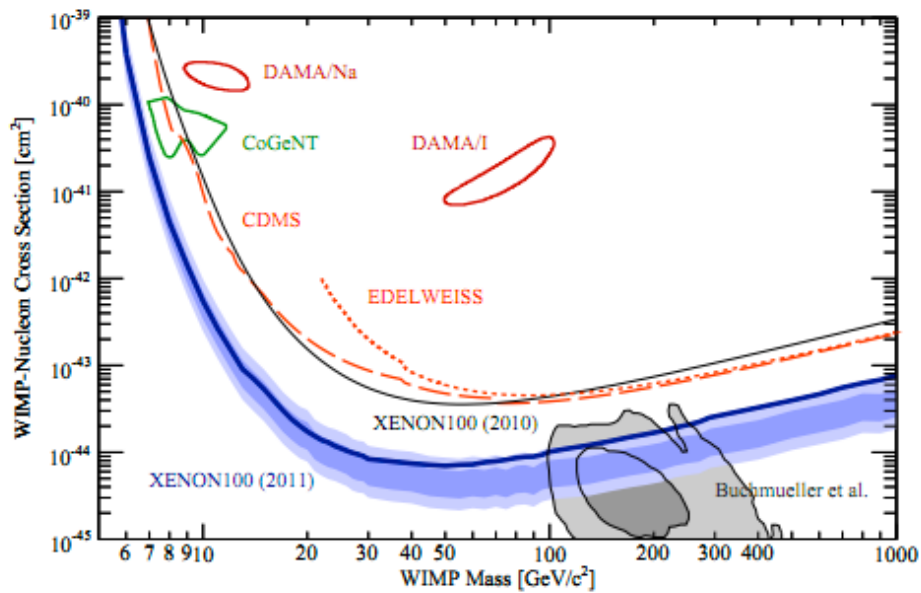

Figure 12. Spin-independent elastic WIMP-nucleon cross-section σ as function of WIMP mass $m_x$. The new XENON100 limit at 90% CL, as derived with the Profile Likelihood method taking into account all relevant systematic uncertainties, is shown as the thick (blue) line together with the 1σ and 2σ sensitivity of this run (shaded blue band). The limits from XENON100 (2010) [7] (thin, black), EDELWEISS [6] (dotted, orange), and CDMS [5] (dashed, orange, recalculated with $v_{esc}$ = 544 km/s, $v_0$ = 220 km/s) are also shown. Expectations from CMSSM are indicated at 68% and 95% CL (shaded gray) [17], as well as the 90% CL areas favored by CoGeNT (green) [18] and DAMA (light red, without channeling).

The new XENON 100 data does not support the DAMA claims (Reference 11).

## 11. The source of annual variation of signals in underground laboratories

It is very well known that many detectors have observed annual variations in nearly all underground laboratories. There are at least three sources:

(1) µ flux modulation
(2) neutron flux modulation (see Figure 3)
(3) Radon abundance modulation

All of these effects tend to give larger signals in the spring and summer months and lower signals in the fall and winter. Here we show the LVD example (Reference 12).

Muon flux from LVD:

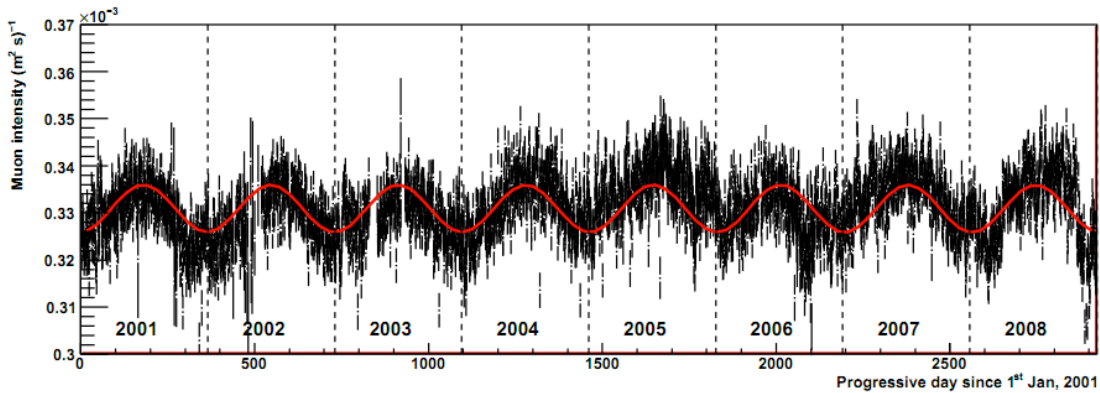

Figure 13. Muon intensity along the 8 years of data acquisition. Each bin corresponds to one day, starting from 1$^{st}$ January, 2001 to 31$^{st}$ December, 2008. The error bars are the statistical uncertainty. The solid red curve is the result of a cosinusoidal fit to the data in the form shown in eq. 3 The vertical dashed lines separate each solar year. Note that the neutrons can be produced underground by either cosmic ray muons or sources like Radon. Therefore the annual variation of the neutrons partially reject the Radon variation. We show again the neutron modulation in Figure 3 that seems to fit the DAMA data.

## 12. The lack of annual variation of the multihit DAMA data

DAMA claims that their multihit events are likely due to neutrons and they display no annual variation. However according to the results of Figure 3 they should follow the same pattern of annual variation of neutrons as the LNGS. So what is the source of these events?
   D. Nygren has suggested that delayed signals in DAMA along with the variation of muon flux from Figure 13 could explain the DAMA effect (Reference 12). The DAMA group strongly denies this conclusion.

## 13. Summary

The result on low-mass WIMPs is confused. Of the three positive signals only DAMA has published results in a peer-reviewed journal claiming direct evidence for dark matter. The negative results come from XENON 100, XENON 10 (two different analyses), and special CDMS runs. All of these results exclude DAMA. There are new direct measurements of $L_{eff}$ and new indirect determinations of $L_{eff}$. This is no longer an issue in these studies. Only the XENON 10 and 100 exclude CoGeNT, CRESST, and DAMA.

At the 2011 DPF meeting at Brown University there was a presentation by the CDMS II group concerning the low mass WIMP region. Using their low noise detectors the exclude the region in Figure 10 left of the DAMA region. Note that the DAMA region in Figure 10 is different from the DAMA region in Figure 7 or Figure 5 for example. Note also in Figure 5 that the outer regions of the DAMA region is unfavored by 35 in the fit in reference 1. Figure 7 also shows a more extreme exclusion of the low mass WIMP region than the new CDMS II results. In the near future we hope there will be new results.

We wish to thank David Nygren for useful discussions and members of the XENON 100 team. We wish to thank the Aspen Center for Physics, where this paper was prepared, for their help.